\documentclass[reprint,superscriptaddress,amsmath,amssymb,aps,prl,floatfix]{revtex4-1}

\usepackage{graphicx}
\usepackage{dcolumn}
\usepackage{bm}
\usepackage{hyperref}

\usepackage{pdfpages}
\usepackage{pgffor}

\newcommand{\npu}{MOE Key Laboratory of Material Physics and Chemistry under Extraordinary Conditions, School of Physical Science and Technology, Northwestern Polytechnical University, Xi’an, 710072, China}
\newcommand{\ecl}{Univ Lyon, Ecole Centrale de Lyon, Laboratoire de Tribologie et Dynamique des Syst\`emes, UMR 5513, 36 avenue Guy de Collongue, 69134 Ecully Cedex, France}
\newcommand{\ilm}{Univ Lyon, Univ Claude Bernard Lyon 1, CNRS, Institut Lumi\`ere Mati\`ere, F-69622, VILLEURBANNE, France}
\newcommand{\iuf}{Institut Universitaire de France (IUF)}

\newcommand{\kt}{k_\text{B}T}

\newcommand{\sigs}{\sigma_\text{s}}
\newcommand{\sigl}{\sigma_\ell}
\newcommand{\sigh}{\sigma_\text{h}}

\newcommand{\lb}{\ell_\text{B}}
\newcommand{\debye}{\lambda_\text{D}}

\makeatletter
\AtBeginDocument{\let\LS@rot\@undefined}
\makeatother

\begin{document}


\title{Liquid-solid slip on charged walls: dramatic impact of charge distribution}

\author{Yanbo Xie}
\thanks{These authors contributed equally to this work.}
\affiliation{\npu}
\author{Li Fu}
\thanks{These authors contributed equally to this work.}
\affiliation{\ecl}
\author{Thomas Niehaus}
\affiliation{\ilm}
\author{Laurent Joly}
\email{laurent.joly@univ-lyon1.fr}
\affiliation{\ilm}
\affiliation{\iuf}

\date{\today}

\begin{abstract}
Nanofluidic systems show great promises for applications in energy conversion, where their performance can be enhanced by nanoscale liquid-solid slip. However, efficiency is also controlled by surface charge, which is known to reduce slip. Combining molecular dynamics simulations and analytical developments, we show the dramatic impact of surface charge distribution on the slip-charge coupling. Homogeneously charged graphene exhibits a very favorable slip-charge relation (rationalized with a new theoretical model correcting some weaknesses of the existing ones), leading to giant electrokinetic energy conversion. In contrast, slip is strongly affected on heterogeneously charged surfaces, due to the viscous drag induced by counter-ions trapped on the surface. In that case slip should depend on the detailed physical chemistry of the interface controlling the fraction of bound ions. Our numerical results and theoretical models provide new fundamental insight on the molecular mechanisms of liquid-solid slip, and practical guidelines for searching new functional interfaces with optimal energy conversion properties, e.g. for blue energy or waste heat harvesting. 
\end{abstract}

\maketitle

\paragraph{Introduction--} 

The development of sustainable alternative energies is one of the greatest challenges faced by our society, and nanofluidic systems 
could contribute significantly in that field \cite{VanderHeyden2006,Pennathur2007,VanDerHeyden2007,Sparreboom2009,Bocquet2014}. 
For instance, membranes with nanoscale porosity could be used to harvest energy from the salinity difference between sea and river water \cite{Siria2013,Feng2016,Siria2017} or from waste heat \cite{Fu2018,Fu2019}. 
Energy conversion in nanofluidic systems originates at liquid-solid interfaces, where the properties of the liquid differ from their bulk value \cite{Anderson1989,Bocquet2010}. 
In particular, in aqueous electrolytes, the so-called electrokinetic (EK) effects --coupling different types of applied forcing and induced flux-- are controlled by 
hydrodynamics and electrostatics in the electrical double layer (EDL), a nanometric charged layer of liquid in contact with charged walls 
\cite{Andelman1995,Hunter2001,Hartkamp2018}. 
Consequently, the EK response of an interface is largely controlled by the wall surface charge \cite{Delgado2007}. Yet, nanoscale liquid-solid slip 
\cite{Bocquet2007,Daivis2018} can amplify EK effects \cite{Marry03bis,Joly2004,Dufreche05bis,Ajdari2006,Ren2008,Maduar2015,Fu2017,Silkina2019,Werkhoven2020}. 
Slip is quantified through the Navier boundary condition (BC), balancing the viscous shear stress at the wall, $\eta\, \partial_z v |_{z=z_w}$ (with $\eta$ the viscosity and $\partial_z v |_{z=z_w}$ the shear rate at the wall), and a liquid-solid friction stress, $\lambda\, v_s$ (with $v_s$ the slip velocity and $\lambda$ the fluid friction coefficient) \cite{Navier1823,Cross2018}. The Navier BC is usually rewritten as: $v_s = b\, \partial_z v |_{z=z_w}$, defining the so-called slip length $b = \eta/\lambda$ \cite{Bocquet2007}. 

In the presence of slip, the EK response is amplified by a factor $1+b/L$, where $L$ is the thickness of the interfacial layer (e.g., the Debye length $\debye$ for the EDL) \cite{Muller1986,Stone2004,Ajdari2006,Bocquet2010}. For optimal performance, it is therefore critical to use surfaces with both a large surface charge and a large slip length. 
With that regard, it has been shown 
that the slip length decreases when surface charge increases \cite{Joly2006,Huang2008,Botan2013,Jing2015}, which impacts the EK energy conversion efficiency \cite{Bakli2015}. 
The slip-charge coupling has been investigated both theoretically and experimentally over the recent years \cite{Jing2013,Pan2013,Li2015b,Catalano2016,Simonnin2018,Mouterde2018,Geng2019}; in particular, a theoretical description has been proposed \cite{Joly2006}
for model surfaces with a homogeneous charge, which can arise from the polarization of a conductive surface, analogous to e.g. the charging of amorphous carbon electrodes in supercapacitors \cite{Pean2015a,Liu2019,Ganfoud2019,Mendez-Morales2019}. 
However, for most surfaces, charge arises from the dissociation of surface groups or specific adsorption of charged species, resulting in a spatially heterogeneous charge. 
Both experiments and simulations have shown that lateral heterogeneity of surface charge can have a strong impact on the interfacial water structure \cite{Cyran2019, Creazzo2019}, and 
in general, it is not clear that the existing theoretical description of slip-charge coupling \cite{Joly2006} is suitable to describe heterogeneous surfaces.

In that context, we used molecular dynamics (MD) simulations to investigate the impact of surface charge distribution on liquid-solid slip, with the goal to understand and optimize the slip-charge dependency. To that aim we considered a model interface between aqueous sodium chloride and charged graphene. We observed a dramatic impact of the surface charge distribution, which we rationalized through analytical modeling. We then explored the consequences of charge distribution on slip-enhanced EK energy conversion, and predicted a giant performance of polarized graphene.

\paragraph{Systems and methods--} 

\begin{figure}
    \centering
    \includegraphics[width=\linewidth]{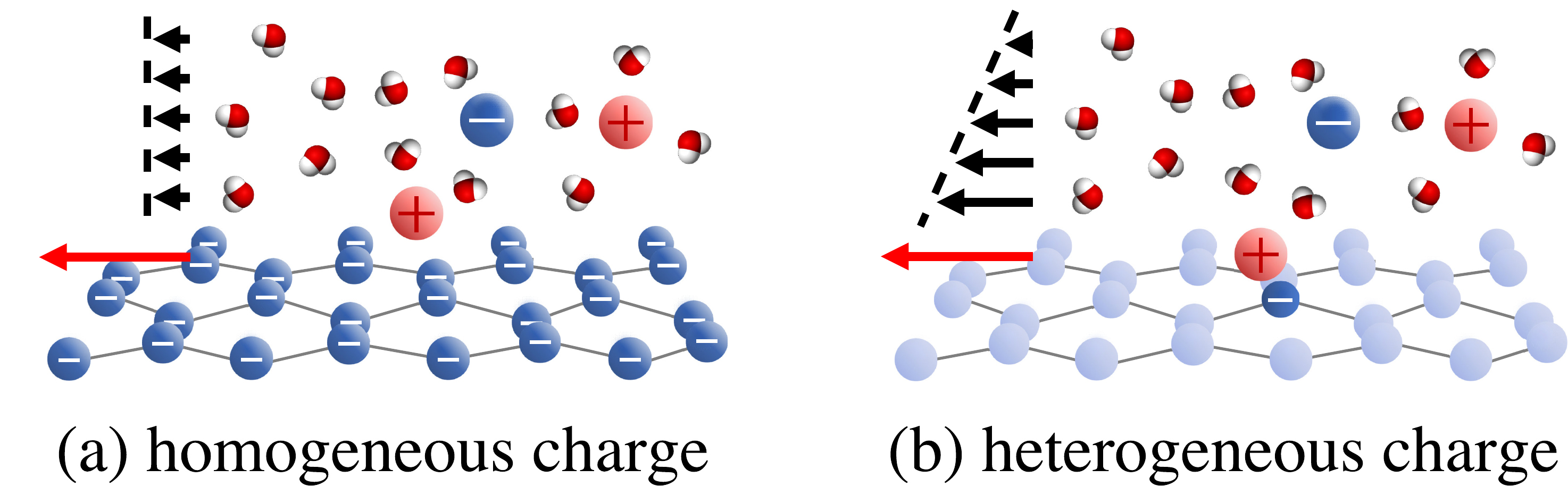}
    \caption{Two surface charge distributions were considered in this work: (a) homogeneous charge, with all surface atoms baring the same partial charge (``polarized wall''); (b) heterogeneous charge, with a fraction of surface atoms baring an elementary charge.} 
    \label{fig:systems}
\end{figure}

We conducted MD simulations with the LAMMPS package \cite{Plimpton1995a} to investigate the change of slip length as function of both the surface charge density and its distribution. 
Here we present the main features of the simulation setup; technical details can be found in the supplemental material (SM) \cite{sm}. 
We considered an aqueous NaCl solution confined between two parallel graphene sheets. 
Previous MD work \cite{Joly2006,Huang2008,Joly2014} has shown that the slip-charge coupling was not significantly affected by the salt concentration, 
and here we used a constant concentration $\rho_s \sim 1.3$\,M in all configurations, unless specified. The corresponding Debye length $\debye$ was ca. 0.26\,nm. 
The distance between the graphene sheets was $\sim 10$ times larger than $\debye$, so that the EDLs of both walls were well separated. We used periodic boundary conditions in the $x$ and $y$ directions parallel to the sheets, with a lateral box size of ca. 3.5\,nm. 
We simulated both homogeneously and heterogeneously charged graphene walls, with surface charge density $\Sigma$ from -0.06 to 0 C/m$^2$ (see Fig.~\ref{fig:systems}). We also considered surfaces with a positive charge, and obtained identical results for homogeneous charge, but different results for heterogeneous charge, as discussed later.   
On homogeneously charged (``polarized'') walls with a surface area of $A$, each atom on a wall was charged by $q/N$, where $q=\Sigma \times A$ is the total charge and $N$ is the total number of carbon atoms on the wall. 
The maximum charge per atom, obtained for $|\Sigma|=0.06$\,mC/m$^2$, was $\sim 0.01 e$. 
We checked using density functional based tight binding (DFTB) \cite{elstner1998scc} simulations that the graphene structure was barely modified by such a charge \cite{sm}. 
On heterogeneously charged walls with the same area and charge density, $n=q/e$ random selected carbon atoms were charged by an elementary charge $e$.

We used the TIP4P/2005 force field \cite{Abascal2005a} for water. Ions were simulated with the scaled-ionic-charge model by \citet{Kann2014}, using a scaling factor of 0.85. 
For consistency, the charge of wall atoms were rescaled with the same factor 
as for the ions \cite{Biriukov2018} in the simulation. 
Nevertheless, we used the unscaled charge for the later calculation of surface charge density. 
Water and carbon interacted through a recently proposed force field calibrated from high-level quantum calculations of water adsorption on graphene \cite{Perez-Hernandez2013}.
The systems were maintained at $T = 298$\,K and $p = 1$\,atm. 
A Couette flow was generated in the liquid by moving the walls with a constant speed of $|V_x|$ in opposite directions along the $x$ axis ($|V_x| = 10-50$\,m/s). 
We employed the same method discussed in Ref.~\citenum{Herrero2019} to compute the slip length \cite{sm}.

\paragraph{Results and discussion--} 

\begin{figure}
    \centering
    \includegraphics[width=0.95\linewidth]{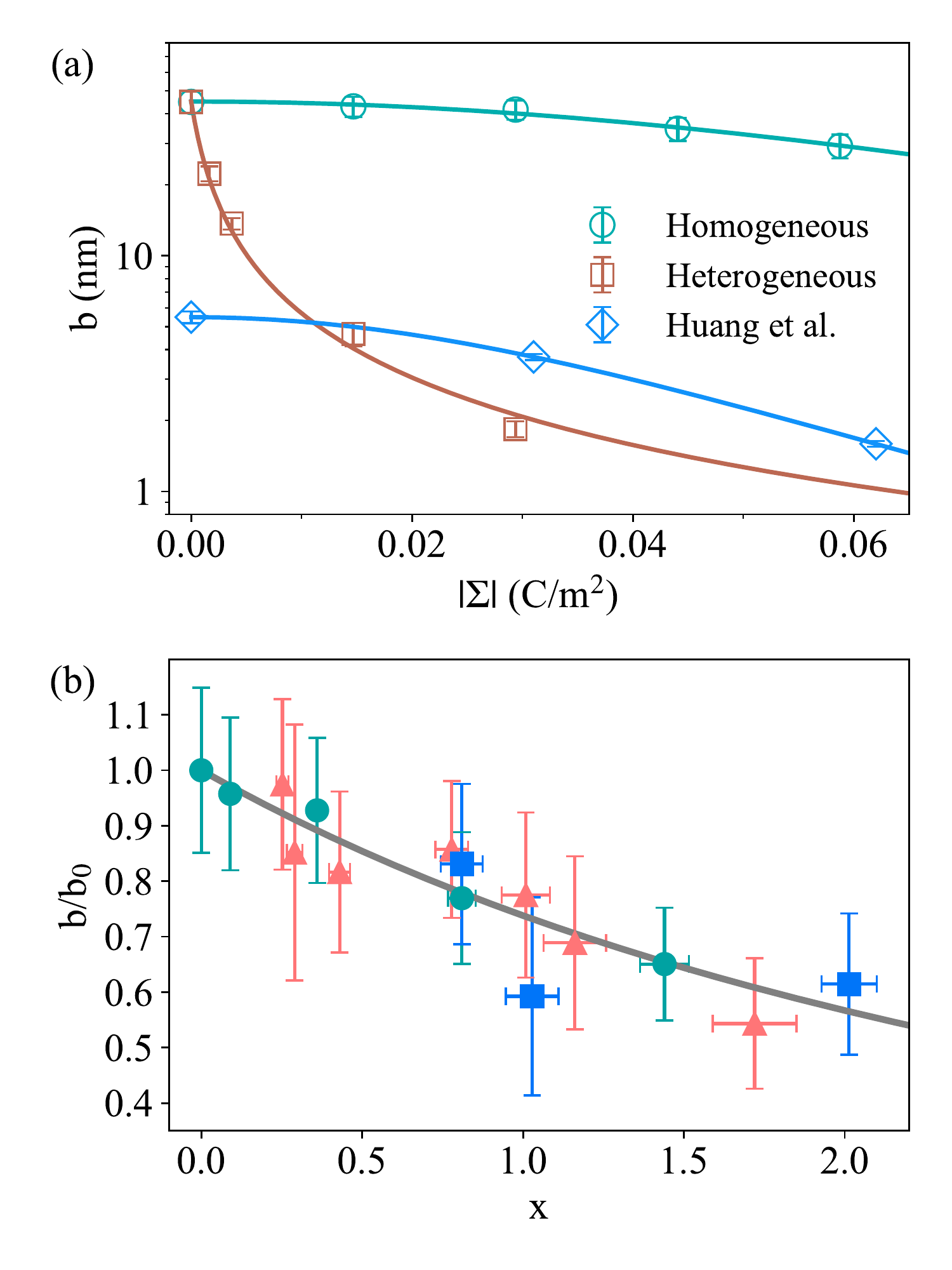}
    \caption{(a) Slip length $b$ versus surface charge density $|\Sigma|$: simulation results for homogeneously (green circles) and heterogeneously (brown squares) charged graphene, and for a generic hydrophobic wall with homogeneous charge (blue diamonds, taken from Ref.~\citenum{Huang2008}); Green and blue lines are fits with Eq.~\eqref{eq:b_homo}, and the brown line is a fit with Eq.~\eqref{eq:b_hete}. 
    (b) Slip length on homogeneously charged surfaces normalized by the uncharged value, $b/b_0$, as a function of the dimensionless parameter $x$ in Eq.~\eqref{eq:b_homo}; simulation results for graphene (green circles) and graphene-like surfaces, either with different wettability (pink triangles) or strained graphene (blue squares); all results are fitted with Eq.~\eqref{eq:b_homo}, using a single value of $\alpha = 0.165$.}
    \label{fig:b-Sigma}
\end{figure}

Figure~\ref{fig:b-Sigma}(a) shows the evolution of $b$ as a function of the surface charge density $|\Sigma|$, for homogeneously and heterogeneously charged graphene walls. For comparison, results from Ref.~\citenum{Huang2008} obtained with a generic hydrophobic surface are also shown. 
Consistently with previous MD results on graphitic surfaces \cite{Kannam2013,Striolo2016}, the slip length on uncharged graphene is very large, ca. 45\,nm. 
Upon charging the surface, the slip length decreases, but the effect of surface charge density on slip is dramatically different between the homogeneous and the heterogeneous walls. On polarized graphene, as $|\Sigma|$ increases from 0 to $0.06$\,C/m$^2$, $b$ gradually decreases from 45 to 30\,nm. On heterogeneously charged graphene, $b$ decays much faster, down by more than a factor of 2 for only $0.015$\,C/m$^2$. 
Finally, comparing the two homogeneous walls, graphene comes out as a more interesting surface than the hydrophobic surface considered in Ref.~\citenum{Huang2008}, combining both a larger slip length on the uncharged wall, and a weaker charge dependency.

In order to rationalize the MD results, and identify criteria for optimal slip-charge dependency, we developed two models to describe the homogeneous and heterogeneous cases. 
For a homogeneous surface charge, we reconsidered a calculation presented in Ref.~\citenum{Joly2006}, as detailed in the SM \cite{sm}. This calculation is based on a Green-Kubo expression for the liquid-solid friction coefficient $\lambda$ (related to the slip length through the viscosity: $b = \eta/\lambda$). The Green-Kubo formula relates $\lambda$ to the fluctuations of the friction force at equilibrium. By separating the electrostatic and the non-electrostatic contributions to the friction force, one can show that: 
\begin{multline}\label{eq:b_homo}
    b = \frac{b_0}{\left(1+\alpha x\right)^2}, \text{with} \\ 
    x = \left( \frac{3\pi \sigl b_0}{\sigs^2} \right)^{1/2} 
    \left( \frac{\lb^{\,i}}{\sigs} \right) 
    \left( \frac{\Sigma \sigs^2}{e} \right)^2 ,
\end{multline}
where $b_0$ is the slip length on the neutral surface, $\alpha$ a numerical prefactor, 
$\sigl$ the effective hydrodynamic diameter of liquid particles, $\sigs$ the wall interatomic distance, and $\lb^{\,i} = e^2 / (4\pi \varepsilon_\text{d}^{\,i} \kt)$ the Bjerrum length of the interface (with $\varepsilon_\text{d}^{\,i}$ the dielectric permittivity of the interface). 
Note that $\alpha$ encompasses the unknown ratio between the corrugation of the tangential electric force and the characteristic normal electric field, which should in particular depends on the crystallographic structure of the wall. 
As discussed in Ref.~\citenum{Huang2008}, because friction arises mainly from interactions between the first liquid adsorption layer and the solid surface, the dielectric permittivity and corresponding Bjerrum length $\lb^{\,i}$ in Eq.~\eqref{eq:b_homo} should be those of the vacuum gap separating these two layers: 
$\lb^{\,i} = \lb^{\,0} \approx 55.8$\,nm at room temperature. 

To fit the MD results for homogeneous charge with Eq.~\eqref{eq:b_homo}, $\sigl$ was obtained from the Stokes-Einstein relation between TIP4P/2005 water self-diffusion and viscosity, characterized in Ref.~\cite{MonteroDeHijes2018}: $\sigl = \kt / (3\pi \eta D) = 0.214$\,nm; $\sigs$ was set to $0.142$\,nm for graphene, and $0.337$\,nm for the generic surface. The only free parameters were therefore $b_0$ and $\alpha$. 
Equation~\eqref{eq:b_homo} fits the MD results very well, using $b_0 = 45.1$\,nm and $\alpha = 0.165$ for graphene, and $b_0 = 5.48$\,nm and $\alpha = 0.270$ for the generic surface. 

In particular, the model shows that the relevant characteristics of the wall controlling the slip-charge dependency are the uncharged slip length (the higher $b_0$ is, the faster $b$ decreases with $\Sigma$) and the wall interatomic distance (the larger $\sigs$ is, the faster $b$ decreases with $\Sigma$). For instance, graphene, having a larger uncharged slip length, should display a stronger slip-charge dependency than the LJ wall, but it benefits from a smaller interatomic distance that overcompensates the effect of the uncharged slip length. Therefore, the behavior of graphene, which combines a large uncharged slip length and a weak slip-charge dependency, can be traced back to the unusually small interatomic distance, and should for that reason be quite unique. Nevertheless, Eq.~\eqref{eq:b_homo} can still be used as a guideline to search for other surfaces with potentially favorable properties.

With that regard, note that Eq.~\eqref{eq:b_homo} differs in several aspects from a similar equation introduced previously, Eq.~(12) in Ref.~\citenum{Joly2006}. First, this new expression does not rely on the assumption that the electric friction is small as compared to the non-electric friction. 
Even in the low surface charge limit, 
the prefactor in front of $\Sigma^2$ scales differently with the uncharged slip length $b_0$: $b_0^{1/2}$ here versus $b_0^1$ in the previous formula. Additionally, the present formula now clarifies how the slip length depends on the liquid and solid atomic sizes.  
To test Eq.~\eqref{eq:b_homo} further and in particular the predicted impact of $b_0$ and $\sigs$, we considered graphene-like surfaces where we varied independently these two parameters. First, we varied the LJ interaction energy between carbon and water atoms without changing the wall structure, in order to change $b_0$ for a constant $\sigs$. 
Second, we considered artificially strained graphene walls, i.e. we changed the inter-atomic distance $\sigs$ 
while keeping the same water-carbon interaction energy as for graphene 
(here $b_0$ was also affected by the strain). 
When doing so we also changed the number of wall unit cells in order to keep the surface (and surface charge density $\Sigma$) approximately constant, and always recomputed the exact value of $\Sigma$ \cite{sm}.  
Figure~\ref{fig:b-Sigma}(b) compares the predictions of the model and the simulation results, which match quite well and validate the new model (details and a comparison with the previous formula are given in the SM).

We now turn to the heterogeneously charged surfaces. 
In that case, counter-ions can strongly bind to the charged sites. In general, the fraction of bound counter-ions should depend on the details of the surface physical chemistry and ion distribution in the EDL. However, in the specific case of the negatively charged graphene surfaces considered in this work, all counter-ions were bound to a charged site, and remained trapped during the whole simulation (we checked that this remained true for a lower salt concentration of $\sim 0.13$\,M). In that case one can consider that the bound counter-ions belong to the solid surface and effectively cancel the surface charge. Consequently, we can estimate the slip length as that of a neutral liquid in contact with a neutral wall, build from the charged wall and the bound counter-ions. The latter protrude over the otherwise smooth surface and generate a Stokes drag, which can be described following a similar derivation used to predict the slip of a liquid over a surfactant layer \cite{Joly2014}. As detailed in the SM \cite{sm}, for monovalent ions one can show that: 
\begin{equation}\label{eq:b_hete}
    b = \frac{b_0}{1 + 3\pi \sigh b_0 (|\Sigma| /e)} ,  
\end{equation}
where $b_0$ is the slip length on the uncharged surface, and $\sigh$ the effective hydrodynamic diameter of the counterions, controlling their individual viscous drag.

Equation~\eqref{eq:b_hete} fits the numerical results very well, using $b_0 = 45.1$\,nm as for the homogeneously charged graphene, and $\sigh = 0.261$\,nm.  
The fitted effective hydrodynamic diameter $\sigh$ of the counter-ions is quite reasonable, with a value close to the Van der Waals diameter of the ions. Furthermore, we show in the SM \cite{sm} that Eq.~\eqref{eq:b_hete} also describes consistently modified graphene with different wettability or interatomic distance. 

However, in general, not all counter-ions will bind to the wall. 
In that case, the slip length will not be directly connected to the surface charge, and will be controlled the fraction of bound ions. Through this fraction, the slip length should therefore depend on the specific physical and chemical features of the interface, in contrast with the homogeneous charge case, where only a few well controlled parameters influence slip. 
As a striking illustration, we simulated heterogeneously charged graphene with a positive charge, see the SM \cite{sm}; in that case, Cl$^-$ counter-ions did not bind to the charged sites, consistently with a previous observation on a similar system by \citet{Qiao2003}, and resulting in a different slip-charge relation.

\begin{figure}
    \centering
    \includegraphics[width=0.9\linewidth]{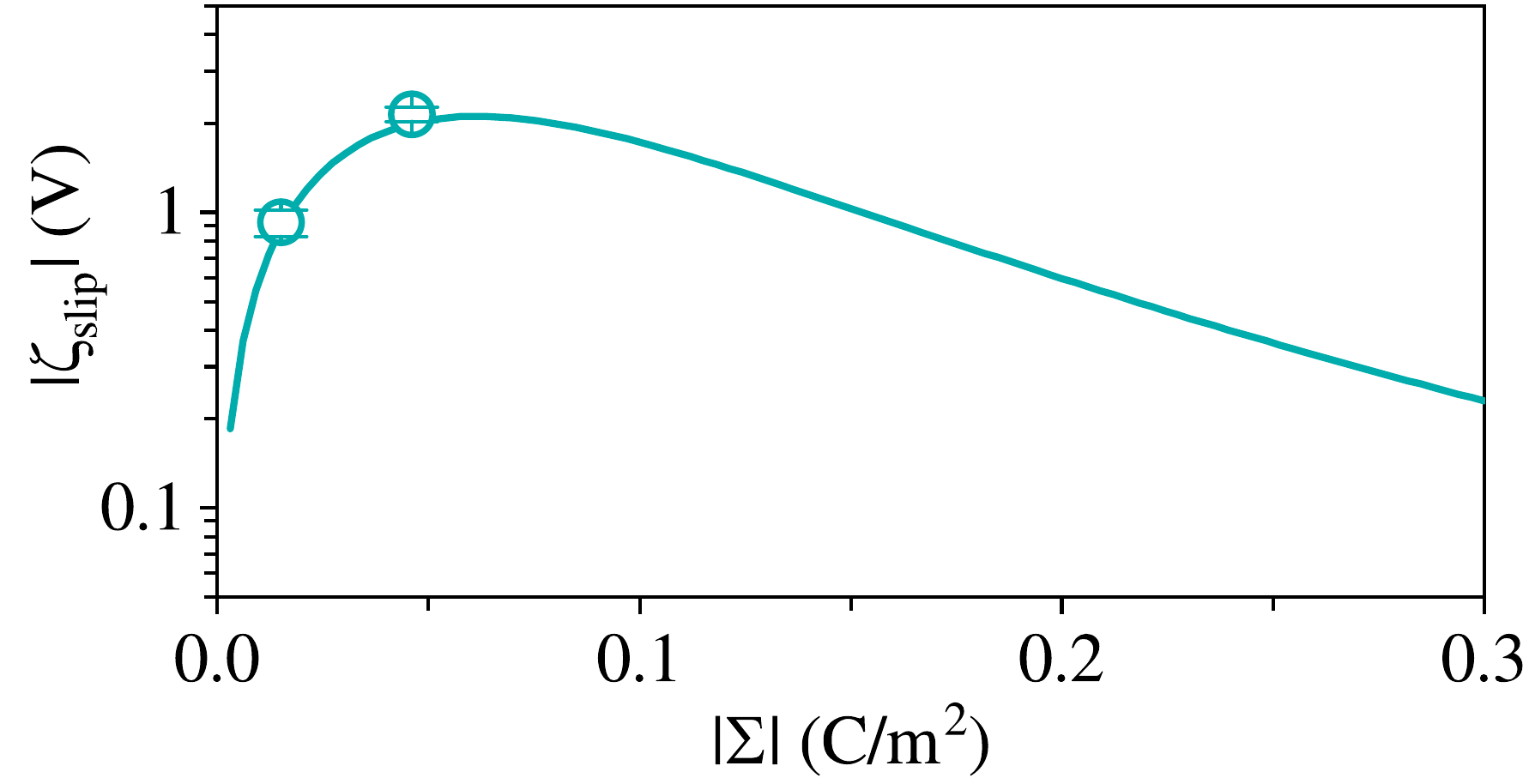}
    \caption{Zeta potential, quantifying the electrokinetic response of the interface, versus surface charge density for homogeneously charged graphene; The line represents the theoretical prediction for the slip contribution $|\zeta_\text{slip}|$; symbols represent measurements through streaming current simulations.} 
    \label{fig:zeta-Sigma}
\end{figure}

We now would like to explore the impact of the slip-charge relation on the energy conversion performance of nanofluidic systems. To that aim, we will focus on electro-mechanical energy conversion at charged surfaces, considering the two reciprocal EK effects of electro-osmotic flows and streaming current \cite{Andelman1995,Hunter2001}. Experimentally, the amplitude of EK effects is quantified by the so-called zeta potential --denoted $\zeta$, extracted from macroscopic measurements of the EK response using the Helmholtz-Smoluchowski (HS) equation \cite{Delgado2007,Hartkamp2018}, which relates the applied forcing and the resulting flux: for electro-osmosis, $v_\text{eo} = -\frac{\varepsilon_\text{d} \zeta}{\eta} E_x$ (with $E_x$ the applied electric field, $v_\text{eo}$ the resulting electro-osmotic velocity, $\varepsilon_\text{d}$ the dielectric permittivity of the liquid), and for streaming current, $j_e = -\frac{\varepsilon_\text{d} \zeta}{\eta} (-\nabla p)$ (with $-\nabla p$ the applied pressure gradient, and $j_e$ the resulting electrical current). 
According to this experimental definition, $\zeta$ is a macroscopic response coefficient, arising from the coupling of electrostatics and hydrodynamics in the EDL. As such, it has been shown theoretically and experimentally that the zeta potential can be amplified by liquid-solid slip \cite{Muller1986,Stone2004,Marry03bis,Joly2004,Bouzigues2008,Audry2010}, and writes \cite{Joly2006}: 
\begin{equation}\label{eq:zeta_slip}
    \zeta = V_0 \left(1 + \frac{b}{\debye^\text{eff}} \right) = V_0 + \frac{\Sigma b}{\varepsilon_\text{d}}, 
\end{equation}
with $V_0$ the surface potential, and where $\debye^\text{eff} = - V_0 / \partial_z V |_{z=z_\text{wall}}$ characterizes the thickness of the EDL. 
The second expression for $\zeta$ shows that liquid-solid slip simply adds a contribution $\zeta_\text{slip} = \Sigma b / \varepsilon_\text{d}$ to the surface potential, which only depends on $\Sigma$, $b$, and $\varepsilon_\text{d}$.

For polarized graphene, using Eq.~\eqref{eq:b_homo} to express $b$,  $|\zeta_\text{slip}|$ is predicted to go through a maximum of $\sim 2000$\,mV, for $|\Sigma| \sim 0.06$\,C/m$^2$ (corresponding to a charge per atom of $\sim 0.01 e$, comparable with charges in the amorphous carbon electrodes of supercapacitors \cite{Pean2015a,Liu2019,Ganfoud2019,Mendez-Morales2019}), see Fig.~\ref{fig:zeta-Sigma} and the SM \cite{sm}. This value exceeds by far usual zeta potentials, which typically saturate around $4\kt/e \sim 100$\,mV. 
To confirm the prediction of the model, we performed explicit streaming current simulations \cite{sm}: 
we applied a pressure gradient to the liquid, measured the resulting electrical current, and computed the zeta potential using the HS equation. The computed zeta potential indeed matches the theoretical prediction, see Fig.~\ref{fig:zeta-Sigma}. 
Consequently, polarized graphene appears as an ideal system to evidence experimentally the zeta potential amplification by liquid-solid slip.

For heterogeneously charged graphene with $\Sigma < 0$, all counter-ions being trapped at the wall, there is no net charge in the liquid, so that the zeta potential must vanish. We performed direct streaming current simulations for $\Sigma = -0.03$\,C/m$^2$ to confirm that prediction and indeed measured a vanishing value within error bars, $\zeta = -3.9 \pm 6.8$\,mV.  

Of course this result is specific to the systems simulated here. For instance, as shown previously \cite{Qiao2003}, simply reversing the surface charge changes the counter-ion adsorption behavior, and consequently the zeta potential. In general, when only a fraction of the counter-ions are trapped \cite{Joly2014,Siboulet2017}, the zeta potential does not vanish, and it depends on the fraction of bound ions, both directly through the resulting effective surface charge and indirectly through the impact of ion binding on slip.

\paragraph{Conclusion--}

Using molecular dynamics simulations and analytical developments, we investigated the impact of surface charge distribution on liquid-solid slip. We focused on model interfaces between aqueous NaCl and graphene. We found a large contrast between surfaces with a homogeneous charge, representative of polarized conductive surfaces, and surfaces with a heterogeneous charge, typically arising from the dissociation of surface groups. 
On polarized graphene, the slip length is very large and weakly affected by surface charge. Our model rationalizes this exceptional performance and traces it back to the unusually small interatomic distance of graphene. 
Note that homogeneously charged graphene was modeled with localized charges, while real polarized graphene features delocalized and mobile charges \cite{CastroNeto2009,Zhan2012}. In future work, ab initio molecular dynamics \cite{Tocci2014a,Grosjean2019,Tocci2020,Mouhat2020} could help to explore the role of electronic screening effects and image charges on liquid-solid friction, for graphene and more generally for metallic walls \cite{Persson2004,Vanossi2013,Sokoloff2018}.  
On heterogeneously charged graphene with a negative surface charge, Na$^+$ counter-ions bind to the charged sites and induce a viscous drag, which strongly decreases the slip length. In contrast, for a positive charge, Cl$^-$ counter-ions do not bind and the slip length decreases less with surface charge. 
Overall, for a heterogeneous surface charge, slip should be affected by the specific details of the ion binding equilibrium, and not be directly controlled by the surface charge, making the development of a generic model for slip-charge coupling particularly challenging.

We also predict a giant EK energy conversion on polarized graphene, due to favorable slip-charge dependency. On heterogeneous surfaces, we predict that the EK response should be specific to the physical chemistry of the interface, both directly through the effective surface charge resulting from counter-ion binding, and indirectly through the impact of bound ions on slip. 
We hope the simulation results and the models developed to rationalize them will help in the search for functional interfaces with optimal EK response. In particular, our results provide a fundamental framework for a future extensive investigation of the complex coupling between ion binding, slip and EK response on a variety of realistic surfaces.

\begin{acknowledgments}
The authors thank Cecilia Herrero, C\'eline Merlet, Mathieu Salanne, and Benjamin Rotenberg for fruitful discussions. 
This work is supported by the ANR, Project ANR-16-CE06-0004-01
NECtAR. YX is supported by NSFC No. U1732143 and Fundamental Research Funds for the Central Universities (Grant No. 3102017jc01001, 3102019ghxm020). LJ is supported by the Institut Universitaire de France. 
\end{acknowledgments}

%



\section*{Supplementary material}

# Supplemental material for "Liquid-solid slip on charged walls: dramatic impact of charge distribution"

Yanbo Xie,[1, *] Li Fu,[2, *] Thomas Niehaus,[3] and Laurent Joly[3, 4, †]

[1]*MOE Key Laboratory of Material Physics and Chemistry under Extraordinary Conditions, School of Physical Science and Technology, Northwestern Polytechnical University, Xi'an, 710072, China*

[2]*Univ Lyon, Ecole Centrale de Lyon, Laboratoire de Tribologie et Dynamique des Systèmes, UMR 5513, 36 avenue Guy de Collongue, 69134 Ecully Cedex, France*

[3]*Univ Lyon, Univ Claude Bernard Lyon 1, CNRS, Institut Lumière Matière, F-69622, VILLEURBANNE, France*

[4]*Institut Universitaire de France (IUF)*

# CONTENTS

## I. SYSTEMS AND METHODS

We computed the slip length and the zeta potential of aqueous electrolyte solutions on graphene surfaces using molecular dynamics (MD). All simulations were performed with the

---

* These authors contributed equally to this work.

† laurent.joly@univ-lyon1.fr

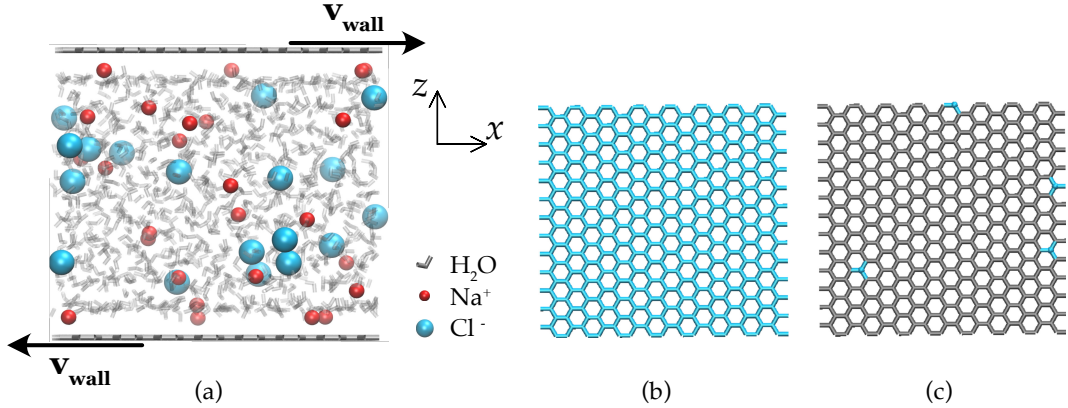

FIG. S1. (a) Snapshot of a typical simulation setup. (b) Top view of the homogeneously charged graphene surface, where all atoms bare a fraction of elementary charge. (c) Top view of the heterogeneously charged graphene surface, where a fraction of atoms bare an elementary charge.

LAMMPS package [S1]. We used Moltemplate [S2] and the VMD [S3] plugin TopoTools [S4] to prepare the initial configurations. The visualization was also realized using VMD [S3].

The system consisted of an aqueous NaCl solution confined between two graphene walls, see Fig. S1. The salt concentration was set to $\rho_s \sim 1.3\,\text{M}$ for all simulations, unless specified. Surface charge densities $\Sigma$ ranged from $-0.06\,\text{C/m}^2$ to $0\,\text{C/m}^2$. We also considered positive surface charges, but we didn't observe a significant impact of the sign of the charge on the slip length.

Unless specified, we considered graphene walls using the experimental structure, with an interatomic distance $\sigma_s = 0.142\,\text{nm}$. Each wall was $\sim 3.4\,\text{nm}$ in length and $\sim 3.2\,\text{nm}$ in width. Note that to test our theoretical description of the slip-charge coupling, we also considered strained graphene, with a modified $\sigma_s$. Strained graphene walls can have a slightly different size to address the requirements of periodic boundary conditions in the lateral $x$ and $y$ directions (more details below). The typical equilibrium distance between these walls was $\sim 3\,\text{nm}$, more than 11 times the Debye length $\lambda_D$ ($\lambda_D \approx 0.26\,\text{nm}$ for TIP4P/2005 water at room temperature and $\rho_s \sim 1.3\,\text{M}$). The typical number of water molecules was $\sim 1000$.

We used the TIP4P/2005 force field for water [S5] and the scaled-ionic-charge model by Kann and Skinner [S6] for ions. As discussed in our previous study [S7], these force fields were found to give good agreement with experimental water diffusion trends and other properties including solution density, radial distribution functions, and ion diffusion

3

coefficients. In this model, a charge of $\pm 0.85\,e$ was used for Na and Cl ions, with $e$ being the elementary charge. Accordingly, each atom on a homogeneously charged wall was charged by $0.85\,q/N$, where $q = \Sigma \times A$ with $A$ being the surface area, and $N$ is the total number of carbon atoms on the wall. On heterogeneously charged walls with the same surface area and charge density, $n = q/e$ random selected carbon atoms were charged by a scaled elementary charge $0.85\,e$. Note that the same treatment has also been used in previous studies on rutile surfaces [S8]. For water-carbon interactions, we took the Lennard-Jones interaction parameters from Ref. [S9], with $\sigma_{\text{CO}} = 0.3157\,\text{nm}$, $\epsilon_{\text{CO}} = 0.5\,\text{kJ/mol}$, $\sigma_{\text{CH}} = 0.2726\,\text{nm}$ and $\epsilon_{\text{CH}} = 0.25\,\text{kJ/mol}$. Finally, carbon-ion interactions were computed consistently with Lorentz-Berthelot mixing rules. In all simulations, we used a cutoff radius of $0.85\,\text{nm}$ for both Lennard-Jones interactions and real-space Coulomb interactions. Long-range Coulombic interactions were computed using the particle-particle particle-mesh (PPPM) method, and water molecules were held rigid using the SHAKE algorithm.

## A. Slip length

We generated a Couette flow to characterize liquid-solid slip. The electrolyte solution was always maintained at a constant temperature of $298\,\text{K}$ using a Nosé-Hoover thermostat, applied only on the degrees of freedom perpendicular to the flow, i.e., in the $y$ and $z$ directions. The wall at the bottom of the simulation box was kept at $z = 0$, while the top one moved as a piston during the first $3\,\text{ns}$. During this period, a pressure of 1 atmosphere was applied homogeneously on the top wall in the $z$ direction, and its motion in the $x$ and $y$ directions was prohibited. After that, the top wall was fixed at its equilibrium position and we ensured systematically that in this stage the bulk solution density always reached the theoretical value under 1 atmosphere. We then moved the walls with a constant speed of $|V_x|$ in opposite directions along the $x$ axis ($|V_x| = 10 - 50$ m/s). A Couette flow was then generated. After a relaxation time of $2\,\text{ns}$, the production period was held for at least $5\,\text{ns}$ for each simulation.

Liquid-solid slip is usually described using the partial slip boundary condition [S10], $v_s = b\,\partial_z v$ (with $v_s$ the slip velocity, $\partial_z v$ the shear rate at the wall, and $b$ the slip length), which was initially written in terms of stress balance by Navier [S11, S12]: at the interface, the viscous shear stress in the liquid, $\eta\,\partial_z v$ (with $\eta$ the shear viscosity), is equal to an

4

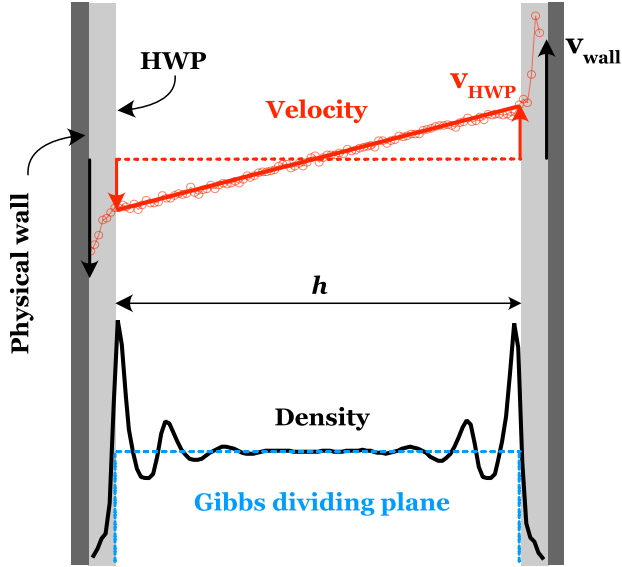

FIG. S2. Illustration of a velocity profile obtained in the simulation and the corresponding liquid density profile. The hydrodynamic wall position (HWP) was determined as the Gibbs dividing plane (GDP). The hydrodynamic height $h$ was computed using the definition of GDP.

interfacial friction stress proportional to the slip velocity: $\tau_w = \lambda\, v_s$ (with $\lambda$ the liquid-solid friction coefficient). Accordingly, the slip length and the friction coefficient are inversely proportional for a fluid with a given viscosity: $b = \eta/\lambda$.

Figure S2 illustrates how we computed the slip length. In a Couette flow, $v_s$ can be obtained as the difference between the imposed wall velocity $v_{\text{wall}}$ and the extrapolated bulk velocity at the hydrodynamic wall position (HWP), i.e., $v_s = v_{\text{wall}} - v_{\text{HWP}}$. In Ref. S13, the authors have shown that the hydrodynamic wall position identifies with the Gibbs dividing plane (GDP), corresponding to a partitioning of space between a region filled with a homogeneous liquid and another one without any liquid. We computed the extrapolated bulk velocity at the hydrodynamic wall position as $v_{\text{HWP}} = \dot{\gamma}\,(h/2)$, with $\dot{\gamma}$ being the bulk shear rate, which is equal to the slope of linear fit of the bulk velocity profile, and $h$ being the hydrodynamic height. According to the definition of the Gibbs dividing plane:

$$h = \frac{N}{n_{\text{bulk}} A}, \tag{S1}$$

where $N$ is the number of liquid particles, and $n_{\text{bulk}}$ is the bulk liquid number density. The friction coefficient $\lambda$ can then be computed as the ratio between the interfacial friction stress $\tau_w$ and the slip velocity $v_s$. Because shear stress is homogeneous along the liquid slab,

5

the bulk viscosity $\eta$ can be obtained as the ratio between $\tau_w$ and the bulk shear rate $\dot{\gamma}$. Finally, the slip length $b$ is obtained as the ratio between $\eta$ and $\lambda$. At least four independent simulations were run for each configuration with different wall velocities. All the results shown in the main text (including the slip length $b$ and the $\zeta$ potential) were obtained by averaging the results which belonged to the linear response regime. The error bars were given as the statistical error within 95% of confidence level.

### B. Zeta potential

To compute the zeta potential, we used a streaming current approach [S14, S15]. In the same systems presented above, we applied an external force $f_i$ on liquid particles in the $x$ direction while keeping the walls immobile, which generates a Poiseuille flow. The pressure gradient $-\nabla p = f_V$ corresponding to this particle force can be computed as $f_V = f_i N/V$ where $N$ is the total number of liquid particles and $V$ is the slit volume. The electric current density $j_e$ was recorded in the simulations, and can be related to the pressure gradient through the Helmholtz–Smoluchowski equation:

$$j_e = -\frac{\varepsilon_\mathrm{d}\zeta}{\eta}\left(-\nabla p\right), \tag{S2}$$

with $\varepsilon_\mathrm{d}$ and $\eta$ the bulk dielectric permittivity and viscosity of the liquid, respectively. The linear fit of $j_e$ against $\nabla p$ gives likewise the $\zeta$ potential, using the bulk values of $\varepsilon_\mathrm{d}$ and $\eta$ for TIP4P/2005 water.

### C. Artificial graphene-like surfaces

As discussed in the main text and mentioned above, we artificially modified the structure of graphene to test our model, Eq. (1) in the main text. First, we varied the LJ interaction energy between carbon and water atoms. This changed the value of $b_0$ while the carbon-carbon distance $\sigma_\mathrm{s}$ was preserved. Second, we varied the distance $\sigma_\mathrm{s}$ from 0.119 to 0.200 nm while keeping all the parameters in the force field unchanged. The neutral slip length $b_0$ was accordingly affected since the wall corrugation had been modified. Due to the requirement of periodic boundary conditions, the lateral lengths of the wall in the $x$ and $y$ directions had to be adjusted, while the surface area was kept roughly constant by changing the number of

TABLE I. Simulation parameters for the graphene-like surfaces.

| | $\epsilon_{\text{CO}}$ (kJ/mol) | $\epsilon_{\text{CH}}$ (kJ/mol) | $\sigma_{\text{s}}$ (nm) |
|---|---|---|---|
| Real graphene | 0.5 | 0.25 | 0.142 |
| LJ interaction modified 1 | 0.22 | 0.11 | 0.142 |
| LJ interaction modified 2 | 0.41 | 0.20 | 0.142 |
| LJ interaction modified 3 | 0.58 | 0.28 | 0.142 |
| LJ interaction modified 4 | 0.87 | 0.43 | 0.142 |
| C-C distance modified 1 | 0.5 | 0.25 | 0.119 |
| C-C distance modified 2 | 0.5 | 0.25 | 0.168 |
| C-C distance modified 3 | 0.5 | 0.25 | 0.200 |

unit cells along the $x$ and $y$ directions. All the modified parameters in this part are presented in table I.

## II. THEORY

Here we derive the expressions for the slip length and zeta potential on homogeneously and heterogeneously charged surfaces.

### A. Homogeneous charge

We first consider the case of a homogeneous surface charge, resulting e.g. from the polarization of a conductive surface. In that case the charge is evenly distributed between all atoms at the wall surface. We will then reconsider the calculation presented in Ref. S14, with some differences, which we will discuss at the end.

#### 1. Slip length

As discussed in Sec. I A, the slip length $b$ can be expressed as the ratio between the bulk viscosity $\eta$ and the interfacial friction coefficient $\lambda$: $b = \eta/\lambda$. The friction coefficient can be obtained from a Green-Kubo formula, i.e. as the time-integral of the autocorrelation

function of the fluctuating friction force $F(t)$ at equilibrium [S16, S17]:

$$\lambda = \frac{1}{Ak_\text{B}T} \int_0^\infty \langle F(t)F(0)\rangle \text{d}t, \tag{S3}$$

with $A$ the interface area, $k_\text{B}$ the Boltzmann constant, and $T$ the temperature. Equation (S3) can be rewritten:

$$\lambda = \frac{\langle F^2\rangle}{A} \times \frac{\tau_\text{F}}{k_\text{B}T}, \tag{S4}$$

where $\langle F^2\rangle$ is the variance of the friction force, and $\tau_\text{F}$ the force decorrelation time defined as $\tau_\text{F} = \int_0^\infty \langle F(t)F(0)\rangle/\langle F^2\rangle \text{d}t$. The latter can be estimated as the diffusive time of liquid molecules over the characteristic spatial period of the wall corrugation $\sigma_\text{s}$, on the order of the wall interatomic distance: $\tau_\text{F} = \sigma_\text{s}^2/D$, with $D$ the liquid self-diffusion coefficient [S10]. The latter can be expressed with an effective Stokes-Einstein law: $D = k_\text{B}T/(3\pi\eta\sigma_\ell)$, with $\sigma_\ell$ the effective hydrodynamic diameter of liquid particles. Overall $\tau_\text{F}$ can then be estimated as:

$$\tau_\text{F} = \frac{3\pi\eta\sigma_\ell\sigma_\text{s}^2}{k_\text{B}T}. \tag{S5}$$

Following Ref. S14, we then separate the electric contribution to the friction force (due to partial charges on wall atoms) from the other contributions. In the simulations, the non-electric contribution arises from Lennard-Jones interactions, and will be denoted $F_\text{LJ}(t)$, but this term generally takes into account all interactions that would remain when the surface charge vanishes. The electric contribution will be denoted $F_\text{ES}(t)$. The variance of the force can be written: $\langle F(t)^2\rangle = \langle [F_\text{LJ}(t) + F_\text{ES}(t)]^2\rangle = \langle F_\text{LJ}(t)^2 + 2F_\text{LJ}(t)F_\text{ES}(t) + F_\text{ES}(t)^2\rangle$. Because the potential energy surfaces from which $F_\text{LJ}(t)$ and $F_\text{ES}(t)$ derive both originate in the same wall atom structure, we assume that the time evolutions of $F_\text{LJ}(t)$ and $F_\text{ES}(t)$ are synchronized, so that one can write:

$$\langle F(t)^2\rangle = F_\text{LJ}^2 + 2F_\text{LJ}F_\text{ES} + F_\text{ES}^2, \tag{S6}$$

where $F_\text{LJ} = \sqrt{\langle F_\text{LJ}(t)^2\rangle}$ and $F_\text{ES} = \sqrt{\langle F_\text{ES}(t)^2\rangle}$.

The LJ contribution $F_\text{LJ}$ can be related to the friction coefficient $\lambda_0$ and to the slip length $b_0$ on a charge-neutral surface. Indeed, in that case, $F_\text{ES} = 0$ and one can write:

$$\lambda_0 = \frac{\eta}{b_0} = \frac{F_\text{LJ}^2}{A} \times \frac{\tau_\text{F}}{k_\text{B}T}. \tag{S7}$$

Combining Eq. (S7) and Eq. (S5), one obtains:

$$F_\text{LJ} = \left(\frac{\sigma_\text{s}^2}{3\pi\sigma_\ell b_0}\right)^{1/2} \times \frac{k_\text{B}T\sqrt{A}}{\sigma_\text{s}^2}. \tag{S8}$$

As detailed in Ref. S14, the electrostatic contribution $F_{\text{ES}}$ should scale like the total charge of the electrical double layer (EDL), $Q \propto \Sigma$, multiplied by the electric field at the interface $E \propto \Sigma/\varepsilon_{\text{d}}^i \propto \Sigma \times \ell_{\text{B}}^i k_{\text{B}} T/e^2$ (with $\Sigma$ the surface charge density, $\varepsilon_{\text{d}}^i$ the dielectric permittivity of the interface, $\ell_{\text{B}}^i$ the corresponding Bjerrum length, and $e$ the elementary charge). Overall one expects that $F_{\text{ES}} \propto (\Sigma/e)^2 \ell_{\text{B}}^i k_{\text{B}} T$. On dimensional grounds, one can go further and write:

$$F_{\text{ES}} = \alpha \left(\frac{\Sigma \sigma_{\text{s}}^2}{e}\right)^2 \left(\frac{\ell_{\text{B}}^i}{\sigma_{\text{s}}}\right) \frac{k_{\text{B}} T \sqrt{A}}{\sigma_{\text{s}}^2}, \tag{S9}$$

with $\alpha$ a numerical prefactor. As discussed in Ref. S18, because friction arises mainly from interactions between the first liquid adsorption layer and the solid surface, the dielectric permittivity and corresponding Bjerrum length $\ell_{\text{B}}^i$ should be those of the vacuum gap separating these two layers. Hence the Bjerrum length will be that of vacuum, $\ell_{\text{B}}^i = \ell_{\text{B}}^0 \approx 55.8\,\text{nm}$ at room temperature.

The friction coefficient $\lambda$ on a charged surface then writes:

$$\lambda = \frac{\eta}{b} = \frac{(F_{\text{LJ}}^2 + 2 F_{\text{LJ}} F_{\text{ES}} + F_{\text{ES}}^2)}{A} \times \frac{\tau_{\text{F}}}{k_{\text{B}} T}. \tag{S10}$$

Combining Eqs. (S5), (S8), (S9), and (S10), one obtains the expression for the slip length on a homogeneously charged surface as a function of the slip length on the corresponding neutral surface $b_0$, the surface charge density $\Sigma$, the hydrodynamic diameter of liquid particles $\sigma_\ell$, and the wall interatomic distance $\sigma_{\text{s}}$:

$$b = \frac{b_0}{(1+\alpha x)^2}, \text{ with } x = \left(\frac{3\pi \sigma_\ell b_0}{\sigma_{\text{s}}^2}\right)^{1/2} \left(\frac{\ell_{\text{B}}^0}{\sigma_{\text{s}}}\right) \left(\frac{\Sigma \sigma_{\text{s}}^2}{e}\right)^2. \tag{S11}$$

Equation (S11) differs in several aspects from a similar equation introduced previously, Eq. (12) in Ref. S14. First, this new expression does not rely on the assumption that the electric friction is small as compared to the non-electric friction. This assumption was indeed limiting the range of surface charge densities where the formula could be applied. Using Eqs. (S8) and (S9), one can write that $F_{\text{ES}} \ll F_{\text{LJ}}$ only if:

$$\Sigma \ll \Sigma_{\text{c}} = \frac{e}{\sigma_{\text{s}}^2} \times \left(\frac{\sigma_{\text{s}}}{\alpha \ell_{\text{B}}^0}\right)^{1/2} \times \left(\frac{\sigma_{\text{s}}^2}{3\pi \sigma_\ell b_0}\right)^{1/4}. \tag{S12}$$

If $\Sigma \ll \Sigma_{\text{c}}$, Eq. (S11) simplifies into:

$$b \approx \frac{b_0}{1 + 2\alpha \left(\frac{3\pi \sigma_\ell b_0}{\sigma_{\text{s}}^2}\right)^{1/2} \left(\frac{\ell_{\text{B}}^0}{\sigma_{\text{s}}}\right) \left(\frac{\Sigma \sigma_{\text{s}}^2}{e}\right)^2}, \tag{S13}$$

9

which is more directly comparable to Eq. (12) in Ref. S14. In particular, we obtain the same scaling with the surface charge and Bjerrum length. In contrast, the prefactor in front of $\Sigma^2$ scales differently with the uncharged slip length $b_0$: here $b_0^{1/2}$ versus $b_0^1$ in the previous formula. Additionnaly, the present formula now clarifies how the slip length depends on the liquid and solid atomic sizes.

2. *Zeta potential*

The zeta potential is a transport coefficient that quantifies electrokinetic effects such as electro-osmosis/phoresis and streaming current [S19–S22]. As such, the zeta potential results from the coupling between electrostatics and hydrodynamics in the EDL. In the presence of slip, the zeta potential can therefore be larger than the surface potential $V_0$ [S23–S28], and writes [S14]:

$$\zeta = V_0 \left(1 + \frac{b}{\lambda_{\mathrm{D}}^{\mathrm{eff}}}\right) = V_0 + \frac{\Sigma b}{\varepsilon_{\mathrm{d}}}, \tag{S14}$$

where $\lambda_{\mathrm{D}}^{\mathrm{eff}} = -V_0/\partial_z V|_{z=z_{\mathrm{wall}}}$ characterizes the thickness of the EDL and identifies with the Debye length $\lambda_{\mathrm{D}}$ in the limit of low surface potential/charge. The second expression for $\zeta$ shows that liquid-solid slip simply adds a contribution to the surface potential, and that this contribution only depends on the surface charge $\Sigma$, the slip length $b$, and the dielectric permittivity of the liquid $\varepsilon_{\mathrm{d}}$.

In the following, we will therefore focus on the slip contribution, $\zeta_{\mathrm{slip}} = \Sigma b/\varepsilon_{\mathrm{d}}$. Using Eq. (S11), one can write the latter as:

$$\zeta_{\mathrm{slip}} = \frac{\Sigma b_0}{\varepsilon_{\mathrm{d}} (1+\alpha x)^2}, \text{ with } x = \left(\frac{3\pi \sigma_\ell b_0}{\sigma_{\mathrm{s}}^2}\right)^{1/2} \left(\frac{\ell_{\mathrm{B}}^0}{\sigma_{\mathrm{s}}}\right) \left(\frac{\Sigma \sigma_{\mathrm{s}}^2}{e}\right)^2. \tag{S15}$$

In the main text, we showed that the slip length for homogeneously charged graphene is well described by Eq. (S11) with the following parameters: $b_0 = 43.5\,\mathrm{nm}$, $\alpha = 0.155$, $\sigma_\ell = 0.214\,\mathrm{nm}$, $\sigma_{\mathrm{s}} = 0.142\,\mathrm{nm}$. We plotted the corresponding prediction for $|\zeta_{\mathrm{slip}}|$ as a function of surface charge in the main text. The slip contribution $|\zeta_{\mathrm{slip}}|$ goes through a maximum for $|\Sigma| \sim 0.06\,\mathrm{C/m^2}$, at a value around $2000\,\mathrm{mV}$. This value exceeds by far usual expected values for the zeta potential, which usually saturates around $4k_{\mathrm{B}}T/e \sim 100\,\mathrm{mV}$. Consequently, charged graphene appears as an ideal system to evidence experimentally the amplification of zeta potential by liquid-solid slip.

### B. Heterogeneous charge

We now consider the case of a heterogeneous surface charge, which corresponds to most mechanisms of surface charge generation on isolating materials, e.g. the dissociation of surface groups. In that case only a fraction of surface atoms bare an elementary charge, with a surface density $n_s$ such that $|\Sigma| = e\, n_s$.

#### 1. Slip length

Here we will focus on the heterogeneously charged graphene walls considered in the main text. On such surfaces, and with monovalent counter-ions, MD simulations show that each charged site traps a counter-ion, which remains stuck over very long times. In that case one can consider that the bound counter-ions belong to the solid surface. Consequently, we will now compute the slip length of a neutral liquid in contact with a neutral wall, build from the charged wall and the bound counter-ions. The latter protrude over the otherwise smooth surface and generate a Stokes drag, which we will describe following a similar derivation used to predict the slip of a liquid over a surfactant layer [S29].

To that aim, we consider a non-equilibrium situation where the liquid slips over the solid wall, with a slip velocity $v_s$. The friction coefficient is then defined as the ratio between the interfacial force per unit area and the slip velocity:

$$\lambda = \frac{F}{A v_s}, \tag{S16}$$

with $F$ the friction force and $A$ the interface area. The friction force can be decomposed into a contribution from the uncharged smooth surface, $F_0$, and a viscous drag $F_v$ due to the protruding counter-ions.

To estimate the viscous drag $F_v$, we will consider a sparse distribution of bound counter-ions, such that the typical distance between ions is much larger than their hydrodynamic diameter $\sigma_\text{h}$, to be properly defined later. This limit corresponds to a surface density of bound counter-ions $n_s \ll 1/\sigma_\text{h}^2$, hence to a surface charge density $|\Sigma| \ll |\Sigma|_\text{max} = e/\sigma_\text{h}^2$. Taking for instance $\sigma_\text{h} \sim 0.3\,\text{nm}$, one obtains a typical order of magnitude for $|\Sigma|_\text{max} \sim 1.8\,\text{C/m}^2$; surface charges observed in experiments are much below this enormous value.

In that limit, one can consider that each counter-ion generates a viscous drag, expected to be proportional to the liquid viscosity and to the slip velocity. Consequently, this viscous

drag can be described by an effective Stokes law, $F_1 = 3\pi\eta\sigma_\mathrm{h} v_s$, defining the effective hydrodynamic diameter of the counterions $\sigma_\mathrm{h}$. The total viscous drag is then:

$$F_v = F_1 \times n_s A = 3\pi\eta\sigma_\mathrm{h} v_s \times \frac{|\Sigma|}{e} A. \tag{S17}$$

One can then compute the friction coefficient:

$$\lambda = \frac{F_0 + F_v}{A v_s} = \lambda_0 + 3\pi\eta\sigma_\mathrm{h} \times \frac{|\Sigma|}{e}, \tag{S18}$$

where $\lambda_0 = F_0/(A v_s)$ is the friction coefficient of the smooth uncharged surface. Correspondingly, the slip length writes:

$$b = \frac{\eta}{\lambda} = \frac{b_0}{1 + 3\pi\sigma_\mathrm{h} b_0 (|\Sigma|/e)}, \tag{S19}$$

where $b_0 = \eta/\lambda_0$ is the slip length on the smooth uncharged surface.

### 2. Zeta potential

For heterogeneously charged graphene, all counter-ions being trapped at the wall, there is no net charge in the liquid, so that the zeta potential must vanish. We performed direct streaming current simulations for $\Sigma = -0.03\,\mathrm{C/m^2}$ to confirm that prediction and indeed measured a vanishing value within error bars, $\zeta = -3.9 \pm 6.8\,\mathrm{mV}$. Of course this result is specific the the model graphene-like surface considered here. In general, only a fraction of the counter-ions are trapped [S30], and the zeta potential does not vanish. In that case the zeta potential will depend on the ion binding equilibrium both directly through the resulting effective surface charge, and indirectly through the impact of ion binding on slip.

## III. COMPARISON BETWEEN THEORY AND SIMULATION RESULTS

### A. Homogeneous charge

Here we will: report the simulation data for homogeneously charged graphene and graphene-like surfaces; detail the procedure we followed to prepare Fig. 2(b) of the main text; show that, while the new formula, Eq. (1) of the main text, captures the charge-friction dependency on all surfaces, the previous formula, Eq. (12) in Ref. S14, fails to describe consistently surfaces with very different uncharged slip lengths $b_0$, and to describe

TABLE II. Homogeneous charge: graphene

| $\varepsilon_{\text{CO}}$ (kJ/mol) | $\sigma_{\text{s}}$ (nm) | $\Sigma$ (mC/m$^2$) | $b$ (nm) | $\Delta b$ (nm) |
|---|---|---|---|---|
| 0.5 | 0.142 | 0 | 44.9 | 4.7 |
| 0.5 | 0.142 | -14.7 | 43.0 | 4.2 |
| 0.5 | 0.142 | -29.4 | 41.7 | 3.9 |
| 0.5 | 0.142 | -44.1 | 34.6 | 3.9 |
| 0.5 | 0.142 | -58.7 | 29.2 | 3.4 |

TABLE III. Homogeneous charge: graphene-like surface with different $\sigma_{\text{s}}$

| $\varepsilon_{\text{CO}}$ (kJ/mol) | $\sigma_{\text{s}}$ (nm) | $\Sigma$ (mC/m$^2$) | $b$ (nm) | $\Delta b$ (nm) |
|---|---|---|---|---|
| 0.5 | 0.119 | 0 | 309.4 | 36.4 |
| 0.5 | 0.119 | -51.2 | 190.2 | 36.0 |
| 0.5 | 0.169 | 0 | 11.9 | 1.9 |
| 0.5 | 0.169 | -51.8 | 9.9 | 0.7 |
| 0.5 | 0.200 | 0 | 10.0 | 1.6 |
| 0.5 | 0.200 | -51.3 | 5.9 | 1.5 |

the high surface charge regime. We also show results of electronic structure calculations for charged graphene. The goal is to estimate how realistic the charge densities discussed in the main article actually are.

1. *Simulation results*

Tables II, III and IV report the simulation data for homogeneously charged graphene, graphene-like surfaces with different $\sigma_{\text{s}}$ and graphene-like surfaces with different $\varepsilon_{\text{CO}}$, respectively.

2. *Preparation of Fig. 2(b) in the main text*

In order to show that Eq. (1) of the main text – corresponding to Eq. (S11) of this text – captures the friction-charge relation on a variety of graphene-like surfaces with different

TABLE IV. Homogeneous charge: graphene-like surface with different $\varepsilon_{\rm CO}$

| $\varepsilon_{\rm CO}$ (kJ/mol) | $\sigma_{\rm s}$ (nm) | $\Sigma$ (mC/m$^2$) | $b$ (nm) | $\Delta b$ (nm) |
|---|---|---|---|---|
| 0.22 | 0.142 | 0 | 209.4 | 42.9 |
| 0.22 | 0.142 | -58.7 | 85.1 | 11.5 |
| 0.41 | 0.142 | 0 | 64.1 | 9.7 |
| 0.41 | 0.142 | -29.4 | 52.3 | 4.8 |
| 0.41 | 0.142 | -58.7 | 34.8 | 5.3 |
| 0.41 | 0.142 | -293.7 | 0.5 | 0.05 |
| 0.58 | 0.142 | 0 | 29.2 | 4.8 |
| 0.58 | 0.142 | -29.4 | 24.8 | 5.3 |
| 0.58 | 0.142 | -58.7 | 20.1 | 3.1 |
| 0.66 | 0.142 | 0 | 22.1 | 3.2 |
| 0.66 | 0.142 | -29.4 | 21.4 | 1.1 |
| 0.66 | 0.142 | -58.7 | 17.0 | 2.0 |
| 0.87 | 0.142 | 0 | 12.1 | 1.7 |
| 0.87 | 0.142 | -58.7 | 11.3 | 0.7 |

interatomic spacing and wetting properties, we plotted in Fig. 2(b) of the main text the evolution of $b/b_0$ as a function of $x$.

To that aim, for each type of surface (i.e., with a given $\sigma_{\rm s}$ and $\varepsilon_{\rm CO}$), we took $b_0$ and its error from the $\Sigma = 0$ simulation, and used it to compute $b/b_0$, $x$ and their respective errors using standard error propagation formulas.

3. *Comparison with the prediction of the previous model*

In Ref. S14, an equation similar to the new Eq. (S11) was introduced to describe slip-charge coupling on homogeneously charged surfaces. We recall here this equation, where we have corrected a missing factor of $\sigma^{-1}$ in the second term of the denominator, and where we have fixed the prefactor to $2\alpha'$ for consistency with our new formulas, Eqs. (S11) and (S13):

$$b = \frac{b_0}{1 + 2\alpha' \left(\frac{b_0}{\sigma_{\rm s}}\right) \left(\frac{\ell_{\rm B}^0}{\sigma_{\rm s}}\right) \left(\frac{\Sigma \sigma_{\rm s}^2}{e}\right)^2}. \quad (S20)$$

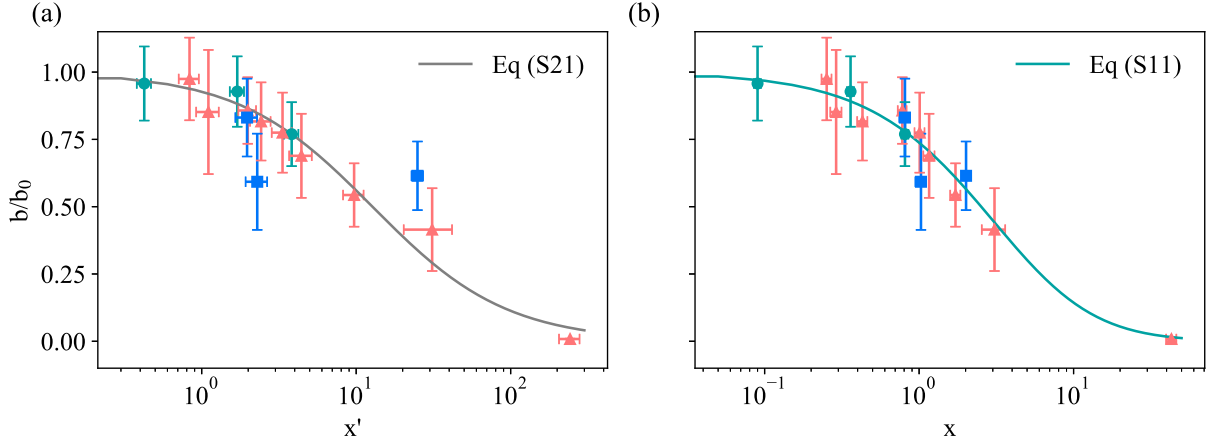

FIG. S3. Simulation results for homogeneously charged graphene (green circles) and graphene-like surfaces, either with different wettability (pink triangles) or strained graphene (blue squares), with the best fit (a) by Eq. (S21) and (b) by Eq. (S11).

This expression relies on the assumption that the electric friction is small as compared to the non-electric friction, and can only be applied for low surface charges. Interestingly, the prefactor in front of $\Sigma^2$ scales differently with the uncharged slip length $b_0$ in Eq. (S20) and in the new Eq. (S11). Consequently, because we have shown that the new formula captured accurately the slip-charge relation on surfaces with very different $b_0$, the previous formula must fail. To illustrate that point, we rewrite Eq. (S20) as:

$$\frac{b}{b_0} = \frac{1}{1 + 2\alpha' x'}, \text{ with } x' = \left(\frac{b_0}{\sigma_s}\right)\left(\frac{\ell_B^0}{\sigma_s}\right)\left(\frac{\Sigma \sigma_s^2}{e}\right)^2. \tag{S21}$$

In the main text we only showed results for moderate surface charges not exceeding $\sim 60\,\text{mC/m}^2$. However, to show that Eq. (S11) also describes well larger surface charges and smaller $b_0$, we ran an additional simulation with $\varepsilon_{\text{CO}} = 0.41\,\text{kJ/mol}$ for $\Sigma \sim 300\,\text{mC/m}^2$ (see table IV). Figure S3 (a) compares the simulation results and the best fit by Eq. (S21) for the evolution of $b/b_0$ as a function of $x'$, using $\alpha' = 0.045$. It is clear that the surfaces with the most extreme values of $b_0$ and/or $\Sigma$ are not captured by this equation. By contrast, the new Eq. (S11) predicts well the results (see Fig. S3 (b)), with the same parameter $\alpha = 0.165$ used for fitting the data in Fig. 1 (b) in the main text.

15

TABLE V. DFTB simulations: Lattice constants of charged graphene in nm. The experimental value was taken from Yazyev and Louie [S33].

| experiment | DFTB (neutral) | DFTB ($+2e$) | DFTB ($-2e$) |
|---|---|---|---|
| 2.46 | 2.4713 | 2.4715 | 2.4727 |

*4. Influence of additional charges on the lattice*

Here we report the results of electronic structure calculations for graphene using the density functional based tight binding method (DFTB) [S31]. DFTB is an approximate DFT method and was extensively used for carbon materials in the past. For the elastic properties of graphene, its accuracy was recently verified in [S32]. We performed periodic DFTB calculations using the *mio-0-1* Slater-Koster parameter set [S31] for a $10 \times 10$ supercell of graphene at the $\Gamma$ point. Relaxation of atomic positions and unit cell with a force treshold of $10^{-4}$ a.u. resulted in a lattice constant of 2.4713 Å. We then charged the $10 \times 10$ cell by $\pm\, 2$ e. This corresponds to a surface charge density of $0.061\,\text{C/m}^2$. The results are given in Table V.

The results show that the additional charge has only a marginal influence on the lattice. For the positively charged graphene the relative change in lattice constant is $+0.01\,\%$, while for the negatively charged graphene it is only slightly larger with $+0.06\,\%$. This shows that graphene is stable for the investigated surface charge density of $0.061\,\text{C/m}^2$.

### B. Heterogeneous charge

Here we will: report the simulation data for heterogeneously charged graphene and graphene-like surfaces; show that Eq. (2) of the main text captures the charge-friction dependency on all surfaces with a negative charge; discuss the qualitatively different ion adsorption behavior on positively charged surfaces and its impact on friction.

*1. Simulation results*

Tables VI, VII and VIII report the simulation data for heterogeneously charged graphene, graphene-like surfaces with different $\sigma_\text{s}$ and graphene-like surfaces with different $\varepsilon_\text{CO}$, respec-

TABLE VI. Heterogeneous charge: graphene

| $\varepsilon_{\rm CO}$ (kJ/mol) | $\sigma_{\rm s}$ (nm) | $\Sigma$ (mC/m²) | $b$ (nm) | $\Delta b$ (nm) |
|---|---|---|---|---|
| 0.5 | 0.142 | 0 | 44.9 | 4.7 |
| 0.5 | 0.142 | -1.6 | 22.3 | 1.6 |
| 0.5 | 0.142 | -3.7 | 13.7 | 0.8 |
| 0.5 | 0.142 | -14.7 | 4.6 | 0.5 |
| 0.5 | 0.142 | -29.4 | 1.8 | 0.1 |
| 0.5 | 0.142 | 29.4 | 3.9 | 0.2 |

TABLE VII. Heterogeneous charge: graphene-like surface with different $\sigma_{\rm s}$

| $\varepsilon_{\rm CO}$ (kJ/mol) | $\sigma_{\rm s}$ (nm) | $\Sigma$ (mC/m²) | $b$ (nm) | $\Delta b$ (nm) |
|---|---|---|---|---|
| 0.5 | 0.119 | -3.2 | 20.8 | 1.6 |
| 0.5 | 0.119 | -12.8 | 6.2 | 1.2 |
| 0.5 | 0.119 | -25.6 | 3.4 | 0.1 |
| 0.5 | 0.168 | -3.2 | 7.9 | 0.4 |
| 0.5 | 0.168 | -13.0 | 3.9 | 0.3 |
| 0.5 | 0.168 | -25.9 | 2.6 | 0.1 |

tively.

2. *Comparison with the model*

To show that our model captures the charge-friction dependency on heterogeneously charged surfaces and graphene-like surfaces, we rewrite Eq. (S19) as:

$$\frac{b}{b_0} = \frac{1}{1 + 3\pi\sigma_{\rm h} y}, \text{ with } y = b_0(|\Sigma|/e), \tag{S22}$$

and we plot the simulation results with the best fit by this formula in Figure S4. It clearly shows that the model depicts well the evolution of $b/b_0$ as a function of $y$, using the same fitting parameter $\sigma_{\rm h} = 0.261$ nm as mentioned in the main text. This effective hydrodynamic diameter $\sigma_{\rm h}$ of the counter-ions is quite reasonable, with a value close to the Van der Waals diameter of the ions.

TABLE VIII. Heterogeneous charge: graphene-like surface with different $\varepsilon_{\mathrm{CO}}$

| $\varepsilon_{\mathrm{CO}}$ (kJ/mol) | $\sigma_{\mathrm{s}}$ (nm) | $\Sigma$ (mC/m$^2$) | $b$ (nm) | $\Delta b$ (nm) |
|---|---|---|---|---|
| 0.41 | 0.142 | -7.3 | 8.6 | 0.4 |
| 0.41 | 0.142 | -14.7 | 4.3 | 0.1 |
| 0.41 | 0.142 | -29.4 | 2.2 | 0.1 |
| 0.41 | 0.142 | -58.7 | 0.9 | 0.1 |
| 0.58 | 0.142 | -7.3 | 6.9 | 0.2 |
| 0.58 | 0.142 | -14.7 | 3.6 | 0.3 |
| 0.58 | 0.142 | -29.4 | 1.8 | 0.2 |
| 0.58 | 0.142 | -58.7 | 0.8 | 0.1 |

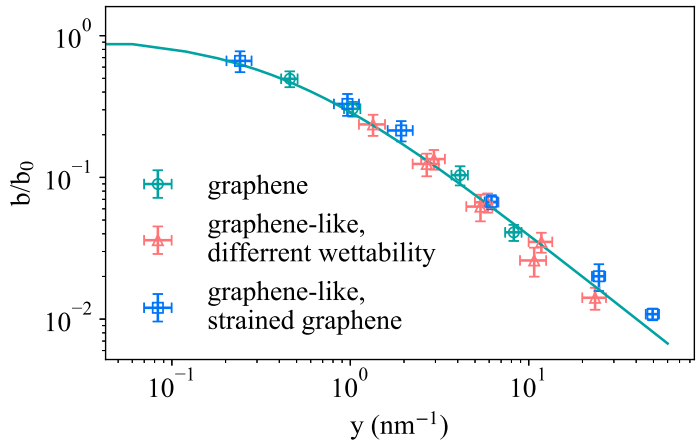

FIG. S4. Simulation results for heterogeneously charged graphene (green circles) and graphene-like surfaces, either with different wettability (pink triangles) or strained graphene (blue squares). The results are fitted by Eq. (S22) using a single value of $\sigma_{\mathrm{h}} = 0.261$ nm.

3. *Ion adsorption on positively charged surfaces*

Previous study by Qiao and Aluru [S34] has shown significant differences between electroosmotic flow of aqueous sodium chloride solutions in positively and negatively charged carbon nanotubes in which the charge was heterogeneously distributed. As noted in the main text, on heterogeneously charged surfaces, the slip length will not only depend on the surface charge, but will also be controlled by the ion binding equilibrium. We performed

18

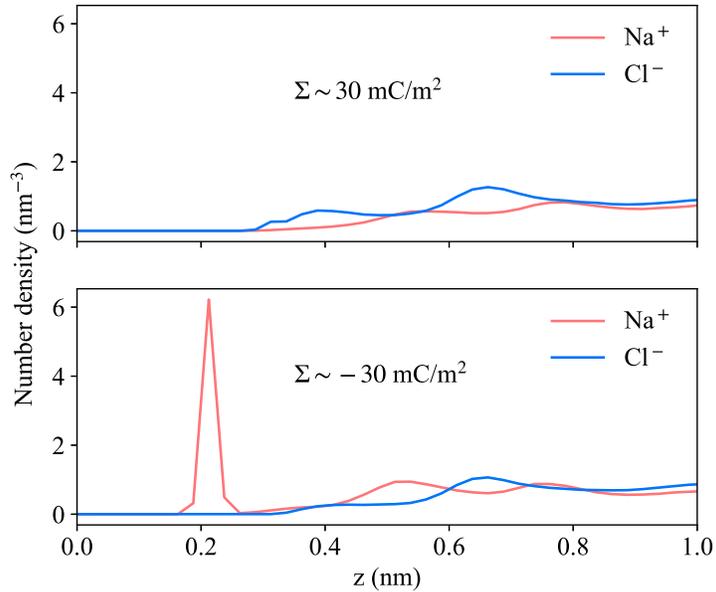

FIG. S5. Density profiles of ions close to the heterogeneously charged walls.

simulations for positively charged surface, with $\Sigma \sim 30\,\text{mC/m}^2$. The result is shown in Table VI and the ion density profiles are plotted in Figure S5 (top). For comparison, those for $\Sigma \sim -30\,\text{mC/m}^2$ are also shown in Figure S5 (bottom). It's clear that for negatively charged graphene, the counter-ions are trapped on the surface, while for positively charged one the counter-ions are layered beyond the wall. Different adsorption equilibrium results in different slip lengths from $\sim 1.8\,\text{nm}$ on negatively charged graphene to $\sim 3.9\,\text{nm}$ on positively charged one. Moreover, there is no effective charge due to the ion binding on negatively charged walls, leading a negligible electroosmotic flow. In contrast, the electroosmotic flow will not vanish on positively charged walls as observed by Qiao and Aluru.



\end{document}